\documentclass[reprint,aps,pra,letterpaper,showpacs,superscriptaddress]{revtex4-1}
\usepackage{amssymb}
\usepackage{amsmath}
\usepackage{bm}
\usepackage{epsfig}
\usepackage{graphicx}
\usepackage[
pdfauthor={Paul D. Nation},
pdftitle={Iterative solutions to the steady state density matrix for optomechanical systems},
bookmarks=true,colorlinks=true,linkcolor=blue,urlcolor=blue,citecolor=red
]{hyperref}

\begin{document}

\title{Iterative solutions to the steady state density matrix for optomechanical systems}

\author{P. D. Nation}
\email{pnation@korea.ac.kr}
\affiliation{Department of Physics, Korea University, Seoul 136-713, Korea}
\author{J. R. Johansson}
\affiliation{iTHES Research Group, RIKEN, Saitama 351-0198, Japan}
\author{M. P. Blencowe}
\author{A. J. Rimberg}
\affiliation{Department of Physics and Astronomy, Dartmouth College, New Hampshire 03755, USA}
\date{\today}

\begin{abstract}
We present a sparse matrix permutation from graph theory that gives stable incomplete Lower-Upper (LU) preconditioners necessary for iterative solutions to the steady state density matrix for quantum optomechanical systems.  This reordering is efficient, adding little overhead to the computation, and results in a marked reduction in \textit{both} memory and runtime requirements compared to other solution methods, with performance gains increasing with system size.  Either of these benchmarks can be tuned via the preconditioner accuracy and solution tolerance.  This reordering optimizes the condition number of the approximate inverse, and is the only method found to be stable at large Hilbert space dimensions.  This allows for steady state solutions to otherwise intractable quantum optomechanical systems.
\end{abstract}


\maketitle
\section{Introduction}\label{sec:intro}
Recently, there has been great interest in generating nonclassical mechanical steady states of optomechanical systems in the single-photon strong coupling regime  \cite{bose:1997,ludwig:2008,rodrigues:2010,rabl:2011,nunnenkamp:2011,kronwald:2013,qian:2012,nation:2013,lorch:2014,lu:2014}, where the standard linearized radiation-pressure approximation breaks down, signifying the ability to generate non-Gaussian steady states of self-oscillating mechanical motion \cite{aspelmeyer:2013, meystre:2013}.  Proposed experimental setups suggest that this regime is now within reach \cite{rimberg:2014, heikkila:2014}, opening up the possibility of observing persistent quantum states of a mechanical oscillator.

The quantum dynamics of an optomechanical system driven by a classical monochromatic pump is given by the Hamiltonian
\begin{equation}\label{eq:hamiltonian}
\frac{\hat{H}}{\hbar}=-\Delta\hat{a}^{\dag}\hat{a}+\omega_{m}\hat{b}^{\dag}\hat{b}+g_{0}(\hat{b}+\hat{b}^{\dag})\hat{a}^{\dag}\hat{a}+E\left(\hat{a}+\hat{a}^{\dag}\right),
\end{equation}
where $\hat{a}$,  $\hat{a}^{\dag}$ and $\hat{b}$,  $\hat{b}^{\dag}$ are annihilation and creation operators for the cavity mode and mechanical resonator, respectively.  Here, we have gone into a frame rotating at the pump frequency $\omega_{p}$, while $\omega_{c}$ and $\omega_{m}$ are the frequency of the cavity and mechanical oscillator modes ($\omega_{c}\gg \omega_{m}$), respectively. The detuning between pump and cavity frequencies is $\Delta=\left(\omega_{p}-\omega_{c}\right)$, while the vacuum radiation pressure coupling strength is denoted by $g_{0}$, and $E$ is the pump amplitude.   We assume that the cavity mode is coupled to the vacuum, with coupling constant $\kappa$, and that the mechanical oscillator is interacting with a possibly nonzero temperature thermal bath with coupling strength $\Gamma_{m}=\omega_{m}/Q_{m}$, where $Q_{m}$ is the mechanical quality factor.

At present, an analytical description of the steady state response of optomechanics is restricted to the linearized regime and the case of single limit-cycle mechanical oscillations with $g_{0}/\kappa \lesssim 1$ \cite{murch:2008, lorch:2014}.  In contrast, the more general case of multiple limit-cycles and single-photon strong coupling,  where the cavity frequency shift per phonon is larger than the cavity line width, $g_{0}/\kappa\gtrsim1$, and a single single photon displaces the mechanical oscillator by more than its zero-point amplitude $g_{0}/\omega_{m}\gtrsim1$ \cite{ludwig:2008,nunnenkamp:2011,aspelmeyer:2013}, has only been explored numerically \cite{qian:2012,nation:2013}.  Importantly, the most nonclassical mechanical states, as measured by negativity of the Wigner function, are known to occur in the presence of multiple limit-cycles  \cite{nation:2013}, thus making numerical analysis of steady state optomechanical systems a critical tool in this regime.  

Ultimately, the simulation of quantum mechanical systems on classical computers is limited by the exponential growth of both memory and runtime of classical methods as the size of the underlying truncated Hilbert space $\mathcal{H}$ increases \cite{feynman:1982, georgescu:2014}, i.e. increasing average photon and phonon occupation number in the steady state.  However, using a sparse matrix representation for quantum operators, the dimensionality of the underlying Hilbert space is currently \textit{not} the limiting factor in determining the steady state solution for a large quantum system.  Rather, it is difficulties brought about by the form of the Liouvillian super operator, describing the quantum dynamics of the optomechanical system interacting with its environment, that are responsible for this limitation.  In particular, it is the lack of Hermicity and poor conditioning that lead to large runtime and memory consumption when using sparse solution methods.  Comprising two coupled, bosonic oscillator modes, optomechanical systems possess a naturally large Hilbert space, and simulations have been restricted to the weak driving limit, and/or that of a heavily damped cavity mode where the average cavity photon number is small.

For large-scale optomechanical systems where, for example, both the cavity and mechanical oscillator have high quality factors, it is commonly necessary to use iterative solvers \cite{saad:2003} that do not perform a lower-upper (LU) factorization of the system Liouvillian.  Instead, this class of solvers uses only matrix-vector multiplication to iteratively converge to an approximate solution vector, and thus are not as memory intensive as direct factorization.  For poorly conditioned systems, incomplete LU (iLU) preconditioners are often used to improve the otherwise slow convergence rate \cite{benzi:2002}.  In general, the condition number grows with the size of the matrix \cite{davis:2006}, and therefore preconditioning is required when solving large ill-conditioned sparse matrices.  When properly preconditioned, these iterative methods converge rapidly, and can even outperform direct solution methods over a wide range of system sizes.  Therefore, using iterative solvers with iLU preconditioning is an important tool for finding the steady state solution to a variety of open quantum systems. Time-dependent methods as applied, for example, in Ref.~\onlinecite{lorch:2014} allow for larger system sizes, but are only applicable to the special case of single mechanical limit-cycles; stochastic switching between multiple limit-cycles restricts the effectiveness of these techniques.

While the generation of robust iLU preconditioners in the case of an Hermitian matrix is now well established, the existence and stability of preconditioners for non-symmetric matrices is understood to a lesser extent \cite{benzi:2002}.  In the non-symmetric case, iLU preconditioners typically fail due to a lack of diagonal dominance, zeros along the diagonal,  and inaccuracies in the approximate inverse due to the dropping of small nonzero elements \cite{chow:1997}.  Moreover, even if a preconditioner is found, its condition number can be larger than that of the original matrix and hence convergence is lost.  However, studies have shown that these failures may be overcome by utilizing symmetric and/or non-symmetric reorderings of the matrix to maximize the sum of the diagonal elements, and reduce the overall bandwidth and profile \cite{chow:1997, benzi:1999, benzi:2002}.  The majority of these reordering strategies are developed for matrices with symmetric structure (graphs) and therefore their application to non-symmetric problems with differing matrix structures must be evaluated on a case by case basis (see Ref.~\cite{benzi:2002} and references therein); there is no universally applicable reordering scheme for non-symmetric matrices.  

To date, the use of reordering methods and preconditioning in solving steady state quantum mechanical systems has yet to be explored.  Iterative methods without reordering or preconditioning were used in Ref.~\onlinecite{lorch:2014}, but the results were limited to small system sizes due to the worsening convergence rate as the size of the system increases.  Furthermore, preconditioning often fails for moderate to large Hilbert space sizes if no reordering strategy is applied to the Liouvillian.  As preconditioned iterative methods are the only means for solving very large-scale linear systems of equations \cite{saad:2003}, finding an appropriate matrix permutation scheme for successful preconditioning is critical to enabling quantum simulations of steady state behavior in systems with large Hilbert space dimension.

Here we show that it is possible to construct robust iLU preconditioners for optomechanical systems by applying the Reverse Cuthill-Mckee (RCM) permutation method \cite{george:1981}, originally developed for graphs from finite-element analysis, to an appropriately symmetrized matrix representation of the system Liouvillian.  This permutation reduces both the bandwidth and profile of the Liouvillian super operator, resulting in a pronounced reduction in memory consumption.  In addition, this reordering increases the stability of the iLU factors by decreasing the condition number of the approximate inverse \cite{chow:1997, bridson:2000, benzi:2002}. This stability in turn allows iterative solvers to converge markedly faster than other steady state solution methods, even when converging to machine precision.  Alternative matrix permutation methods are found to give unstable iLU factorizations, and RCM reordering is the only known method that is stable over the entire range of system parameters.  Trade-offs between memory utilization and runtime can be selectively tuned via the appropriate choice of preconditioning parameters and solution tolerance.  The relative benefit from using this method increases with Hilbert space dimensionality, thus allowing for finding the steady state solution to previously unmanageable high-dimensional optomechanical systems.

This paper is organized as follows.  In Sec.~\ref{sec:methods} we discuss solution methods for the steady state density matrix of an arbitrary Liouvillian super operator represented by a sparse matrix, and show how one is naturally lead to the use of preconditioned iterative methods for large quantum systems.  Section~\ref{sec:reordering} details the reordering strategy used in the case of an optomechanical system, and how this method improves the factorization properties of the Liouvillian.  In Sec.~\ref{sec:numerics} we show the benefit of this reordering with respect to both memory consumption and runtime, as well as the robustness under parameter variations, via numerical simulations.  In addition, we apply this method to a superconducting optomechanical realization \cite{rimberg:2014} and show that persistent nonclassical mechanical states exist for experimentally achievable device parameters.  Finally, Sec.~\ref{sec:conclusion} ends with a discussion of the results, and the application of these methods to other open quantum systems.

\section{Steady state solution methods}\label{sec:methods}
For open quantum systems with decay rates larger than the corresponding excitation rates, the system approaches the steady state density matrix $\hat{\rho}_{\rm ss}$ as $t\rightarrow \infty$ satisfying the equation
\begin{equation}\label{eq:ss}
\frac{d \hat{\rho}_{\rm ss}}{dt}=\mathcal{L}\left[\hat{\rho}_{\rm ss}\right]=0,
\end{equation}
where $\mathcal{L}$ is the Liouvillian super operator, here assumed to be in Lindblad form
\begin{align}\label{eq:lindblad}
\mathcal{L}[\hat{\rho}]=&-i[\hat{H},\hat{\rho}]+\kappa \mathcal{D}\left[\hat{a},\hat{\rho}\right]\\
&+\Gamma_{m}(1+n_{\rm{th}})\mathcal{D}[\hat{b},\hat{\rho}]+\Gamma_{m}n_{\rm th}\mathcal{D}[\hat{b}^{\dag},\hat{\rho}], \nonumber
\end{align}
where the dissipative terms are given by $\mathcal{D}[\hat{O},\hat{\rho}]=\frac{1}{2}[2\hat{O}\hat{\rho}\hat{O}^{\dag}-\hat{\rho}\hat{O}^{\dag}\hat{O}-\hat{O}^{\dag}\hat{O}\hat{\rho}]$, and $n_{\rm th}=[\exp(\hbar\omega_{m}/k_{\rm B}T)-1]^{-1}$ is the average occupation number of the mechanical oscillator's thermal environment at temperature $T$.

If the system Hamiltonian and collapse operators are time-independent,  then Eq.~(\ref{eq:ss}) can be cast as an eigenvalue equation:
\begin{equation}\label{eq:eigval}
\bm{\mathcal{L}}\vec{\rho}_{\rm ss}=0\vec{\rho}_{\rm ss},
\end{equation}
where $\vec{\rho}_{\rm ss}$ is the dense vector formed by vectorization (column stacking) of $\hat{\rho}_{\rm ss}$, and $\bm{\mathcal{L}}$ is the sparse matrix representation of the Liouvillian in a chosen basis.  In what follows, we will assume a Fock state basis representation for all operators.

Equation~(\ref{eq:eigval}) can be solved iteratively as a sparse eigenvalue problem \footnote{Since any eigenvalue problem is equivalent to finding the roots of the characteristic polynomial, it is impossible to directly compute the eigenvalues for any matrix with dimension $\ge5$ \cite{trefethen:1997}.} using, for example, ARPACK \cite{arpack:1997} or an inverse-power method \cite{trefethen:1997} employing a shifted-inverse technique to solve the related system of equations $(\bm{\mathcal{L}}-\sigma \mathbb{I})$, where $\mathbb{I}$ is the identity matrix and $\sigma$ is the requested eigenvalue which is now the dominant eigenvalue of the system \footnote{For zero eigenvalues, it is typical to shift the eigenvalue equation by a small nonzero value, for example $\sigma=10^{-15}$.}.   Unlike Hermitian matrices, where the eigenvalue spectrum is stable against perturbations \cite{trefethen:1997}, the eigenvalues of non-symmetric matrices can be arbitrarily ill-conditioned \cite{saad:1989}, and the associated loss of accuracy can give rise to considerable errors in the computed eigenvalues.  This may be overcome, in part, via eigenvalue balancing \cite{chen:2000}.

 Alternatively, it is possible to find a direct (i.e non-iterative) solution to $\vec{\rho}_{\rm ss}$ via sparse LU decomposition \cite{davis:2006} by making use of the unit trace property of the density matrix to write
\begin{equation}\label{eq:linear}
\tilde{\bm{\mathcal{L}}}\vec{\rho}_{\rm ss}=\left[\bm{\mathcal{L}}+w\bm{\mathcal{T}}\right]\vec{\rho}_{\rm ss}=\left(\begin{array}{c}w \\ 0 \\ \vdots\end{array}\right); \ \ \bm{\mathcal{T}}\vec{\rho}_{\rm ss}=\left(\begin{array}{c}\mathrm{Tr}\hat{\rho}_{\rm ss} \\ 0 \\ \vdots\end{array}\right),
\end{equation}
where $w$ is an arbitrary weighting factor and $\bm{\mathcal{T}}$ is a matrix with ones along the upper row in the columns corresponding to the locations of the diagonal elements in $\vec{\rho}_{\rm ss}$.  Importantly, $\bm{\mathcal{T}}$ always has a nonzero element in the final column.  Unlike iterative methods, the condition number has no effect on direct factorization, making this method attractive for finding steady state solutions to open quantum systems. Note that the restriction $\mathrm{Tr}\hat{\rho}=1$ is already included in the Liouvillian operator $\mathcal{L}$, and its matrix representation $\bm{\mathcal{L}}$.  Here, this constraint is used to simply add a constant vector to both sides of Eq.~(\ref{eq:eigval}).

When performing the LU factorization of a sparse matrix, new nonzero elements arise in the $L$ and $U$ factors; the sparse structure of $L+U$ is not the same as $\tilde{\bm{\mathcal{L}}}$ \cite{davis:2006}.  This fill-in, must be minimized in order to reduce both the memory requirements for storing the LU factors and the runtime of factorization, both of which scale with the number of nonzero matrix elements $\rm NNZ$ \cite{davis:2006}.  The fill-in is sensitive to the order in which the rows and columns of a sparse matrix are operated on, and in particular, to the matrix bandwidth size and profile \cite{george:1981}.  For a non-symmetric sparse matrix $\bm{A}=\{a_{ij}\}$, one can define the upper and lower bandwidths, `$\rm ub$' and `$\rm lb$' respectively, to be
\begin{equation}
\rm{ub}=\max_{a_{ij}\neq0}(j-i); \ \ \rm{lb}=\max_{a_{ij}\neq0}(i-j).
\end{equation}
The total bandwidth $B$ is then the sum $B=\rm{ub}+\rm{lb}+1$, where the one takes into account the main diagonal \cite{saad:2003}.  Likewise, we define the upper profile $\rm up$ and lower profile $\rm lp$ as
\begin{equation}
\rm{up}=\sum_{i} \max_{a_{ij}\neq0}(j-i); \ \ lp=\sum_{j} \max_{a_{ij}\neq0}(i-j),
\end{equation}
such that the total profile is $P=\rm{up}+\rm{lp}$.  Finding a permutation of rows and columns of $\tilde{\bm{\mathcal{L}}}$ that simultaneously reduces both the bandwidth and the profile will be the goal of our reordering strategy.  Iterative eigenvalue algorithms also use LU factorization, and the buildup of fill-in likewise affects these methods.

Unfortunately, direct LU decomposition scales poorly with matrix size in terms of both runtime and memory requirements, even when reordering methods are employed.  Therefore, for sparse matrices of considerable size, iterative methods are the only available method \cite{benzi:2002}, with the most common choice being iterative Krylov solvers \cite{saad:2003}.  While iterative methods require less memory and fewer numerical operations than direct methods, these methods often require preconditioning to achieve a reasonable convergence rate \cite{benzi:2002}.  The goal of preconditioning is to convert the original system of equations, Eq.~(\ref{eq:linear}), into a modified linear system 
\begin{equation}\label{eq:modified}
\bm{\mathcal{M}}^{-1}\tilde{\bm{\mathcal{L}}}\vec{\rho}_{ss}=\bm{\mathcal{M}}^{-1}\left(\begin{array}{c}w \\ 0 \\ \vdots\end{array}\right),
\end{equation}
where $\bm{\mathcal{M}}$ is the preconditoner.  Convergence is improved provided that the condition number of $\bm{\mathcal{M}}^{-1}\tilde{\bm{\mathcal{L}}}$ is significantly lower than that of $\tilde{\bm{\mathcal{L}}}$ itself.  The best preconditioner is obviously $\bm{\mathcal{M}}=\tilde{\bm{\mathcal{L}}}$, however this is equivalent to solving the original system of equations.  Instead, it is possible to efficiently solve for an approximation of the  inverse $\bm{\mathcal{M}}\approx\tilde{\bm{\mathcal{L}}}$.  The application of a suitable preconditioner should make the modified linear system Eq.~(\ref{eq:modified}) easy to solve, and the preconditioner itself should be simple to build and apply as one or more matrix-vector products are required for each iteration.  Moreover, the condition number of $\bm{\mathcal{M}}$ should not be so large as to affect convergence.
\begin{figure*}[t]
\includegraphics[width=17.2cm]{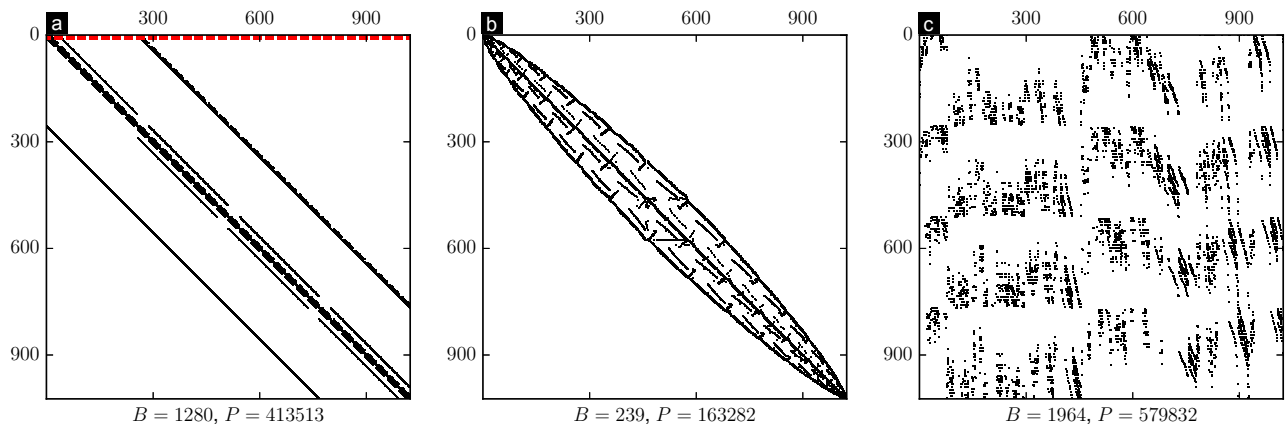}
\caption{(Color online) (a) Sparse matrix structure for the modified Liouvillian $\tilde{\bm{\mathcal{L}}}$ of an optomechanical system with $N_{c}=4$ and $N_{m}=8$ along with the matrix bandwidth $B$ and profile $P$.  Here it is assumed that the mechanical oscillator is in a nonzero thermal environment.   The elements corresponding to matrix $\bm{\mathcal{T}}$ (red), here enlarged for visibility, responsible for the large total bandwidth, are along the upper row.  (b) RCM reordering $\tilde{\bm{\mathcal{L}}}_{\rm RCM}$, where the permuted matrix has a total bandwidth $\sim 5$ times lower than the naturally ordered matrix, and a profile reduction of $\sim 2.5$.  (c) Non-symmetric COLAMD ordering of $\tilde{\bm{\mathcal{L}}}$.  Only the structure of $\tilde{\bm{\mathcal{L}}}$ is required for these reorderings.  The number of nonzero elements is $\rm{NNZ}(\tilde{\bm{\mathcal{L}}})=8957$.}
\label{fig:fig1}
\end{figure*}

The iLU class of preconditoners are constructed from an incomplete (approximate) LU factorization to the modified Liouvillian $\tilde{\bm{\mathcal{L}}}$ by discarding fill-in elements based on a designated dropping strategy.  The method used here is an incomplete LU factorization with dual-threshold and pivoting (iLUTP) \cite{saad:2003}, where a drop tolerance $d$ and allowed fill-in $p$ are specified such that all fill-in smaller than $d$ times the infinity-norm of a row are dropped, and at most only $p\cdot \rm{NNZ}(\tilde{\bm{\mathcal{L}}})$ fill-in elements are allowed.  Note that these parameters are not independent.  In the limit where $d=0$, the iLU preconditioner returns the complete LU factorization, and the fill-in for the direct factorization can be viewed as an upper-bound on the size of the preconditioner.  The condition number of $\bm{\mathcal{M}}$ can vary as a function of the drop tolerance, and therefore decreasing $d$ does not necessarily improve the convergence rate \cite{chow:1997}.  Finding the best combination of parameters is a trial and error process, thus preventing the use of iLU factorization as a ``black box" solver.

\section{Reordering Strategy}\label{sec:reordering}
Finding the minimal bandwidth of a matrix is an NP-complete problem \cite{yannakakis:1981}, and therefore several heuristics have been developed that attempt to minimize the bandwidth while remaining computationally efficient.  One such technique is to permute the rows and columns of a symmetric matrix based on the Cuthill-McKee (CM) ordering \cite{cuthill:1969}.  Taking the structure of a symmetric sparse matrix as the adjacency matrix of a graph, CM ordering does a breadth-first search of the graph starting with a node (row) of lowest degree, where the degree of the $i^{th}$ node is defined to be the number of nonzero elements in the $i^{th}$ matrix row, and visiting neighboring nodes in each level-set in order from lowest to highest degree.  This is repeated for each connected component of the graph.  CM ordering also reduces the profile of the matrix, and it was noticed that reversing the CM order, the RCM ordering, gives a superior profile reduction while leaving the bandwidth unchanged \cite{george:1981}.  Since RCM operates on the structure of a matrix, only the locations of nonzero matrix elements, and not their numerical values, are used in this reordering.  In the Fock basis, the effect of RCM reordering is to permute the basis vectors such that the Fock states are no longer in ascending order.  As the Liouvillian operator is itself non-symmetric, we  calculate the RCM ordering of the symmetrized form $\tilde{\bm{\mathcal{L}}}+\tilde{\bm{\mathcal{L}}}^{T}$ and apply the resulting row and column ordering to $\tilde{\bm{\mathcal{L}}}$ to obtain $\tilde{\bm{\mathcal{L}}}_{\rm RCM}$ \footnote{One can also used the symmetrized product $\tilde{\bm{\mathcal{L}}}\tilde{\bm{\mathcal{L}}}^{T}$ or $\tilde{\bm{\mathcal{L}}}^{T}\tilde{\bm{\mathcal{L}}}$, however this would lead to dense matrices and large memory consumption.}.   

In Fig.~\ref{fig:fig1} we demonstrate RCM reordering on an optomechanical system where the mechanical resonator is coupled to a nonzero thermal bath.  There is an ambiguity when building the sparse matrix representation of the Liouvillian since the tensor product, and therefore matrix structure, depends on the ordering of the operators involved.  For concreteness we take $\hat{a}\equiv \hat{a}\otimes\mathbb{I}_{b}$ and $\hat{b}\equiv \mathbb{I}_{a}\otimes\hat{b}$, although the choice of operator ordering does not affect the results presented here.  From Fig.~\ref{fig:fig1}a we see that imposing the trace condition with the matrix $\bm{\mathcal{T}}$ gives an upper bandwidth that is equal to the square of the dimensionality of the optomechanical Hilbert space, $\mathrm{dim}~\mathcal{H}=N_{c}N_{m}$, where $N_{c}$ and $N_{m}$ are the number of cavity and resonator states, respectively, in the truncated Hilbert space.  Therefore, the total bandwidth must satisfy $B(\tilde{\bm{\mathcal{L}}})>(N_{c}N_{m})^{2}$, suggesting that the fill-in for the LU factors rises rapidly with system size.  Arising from the use of the trace matrix $\bm{\mathcal{T}}$, and not the form of the Liouvillian, this bandwidth scaling holds for any system Hamiltonian.  Applying RCM reordering to $\tilde{\bm{\mathcal{L}}}$ (Fig.~\ref{fig:fig1}b) significantly reduces both the bandwidth and profile of the original matrix.  

As the reduction of fill-in is a well known problem, we are also interested in comparing RCM reordering of $\tilde{\bm{\mathcal{L}}}$ against the best general purpose fill-in reducing permutation for non-symmetric sparse matrices, the Column Approximate Minimum Degree (COLAMD) ordering \cite{davis:2004}.  COLAMD is the default ordering in the SuperLU library  \cite{demmel:1999} used here, as well as in commercial software such as Matlab \cite{matlab}.  The application of this non-symmetric (column only) reordering to the modified Liouvillian is presented in Fig.~\ref{fig:fig1}c.

RCM reordering of $\tilde{\bm{\mathcal{L}}}+\tilde{\bm{\mathcal{L}}}^{T}$ overestimates the graph structure of $\tilde{\bm{\mathcal{L}}}$, and there is little a priori information from which to judge whether this reordering strategy will be successful when the structure of the Liouvillian operator is varied.  Structural changes can occur when setting to zero numerical system parameters in the Hamiltonian (\ref{eq:hamiltonian}) or collapse operators in Eq.~(\ref{eq:lindblad}), as well as when the number of Fock states of the cavity $N_{c}$ and/or mechanical resonator $N_{m}$ are altered.  
\begin{figure}[t]
\includegraphics[width=8.6cm]{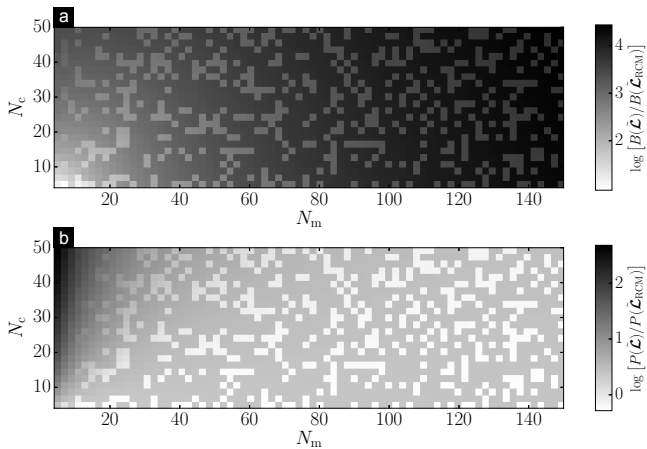}
\caption{(a) Log-plot of the bandwidth reduction factor after RCM reordering for an optomechanical system with mechanical oscillator coupled to a nonzero thermal environment for a varying number of cavity and mechanical resonator states, $N_{c}$ and $N_{m}$ respectively.  (b) Log-plot of the corresponding profile reduction factor.}
\label{fig:fig2}
\end{figure}
Leaving the discussion of numerical coefficients to Sec.~\ref{sec:numerics}, in Fig.~\ref{fig:fig2} we plot the total bandwidth and profile reduction factors, $B(\tilde{\bm{\mathcal{L}}})/B(\tilde{\bm{\mathcal{L}}}_{\rm RCM})$ and $P(\tilde{\bm{\mathcal{L}}})/P(\tilde{\bm{\mathcal{L}}}_{\rm RCM})$, respectively, as the number of optical cavity and mechanical resonator states is varied.  Here it is seen that the RCM method in general reduces both the bandwidth and the profile of $\tilde{\bm{\mathcal{L}}}$, with the benefits of reordering increasing with the state space dimensionality.  The bandwidth and profile of $\tilde{\bm{\mathcal{L}}}_{\rm RCM}$ are not only lower than $\tilde{\bm{\mathcal{L}}}$, but also less than that of $\bm{\mathcal{L}}$ itself.  For some Hilbert space dimensions, RCM reordering does not do as well as expected due to the over estimated graph structure.  In particular, the profile of the reordered matrix can be larger than the original.  However, the RCM ordering always gives a markedly lower bandwidth than the naturally ordered $\tilde{\bm{\mathcal{L}}}$.  For even the largest Hilbert space in Fig.~\ref{fig:fig2}, with $\rm {NNZ}(\tilde{\bm{\mathcal{L}}})\sim 7\times 10^{8}$, RCM reordering takes less than one minute on a $2.3~\rm GHz$ CPU.  As this is a fraction of the total steady state computation time, one can efficiently examine the bandwidth and profile reduction of RCM ordering before performing factorization.

\section{Numerical simulations}\label{sec:numerics}
\begin{figure}[b]
\includegraphics[width=8.6cm]{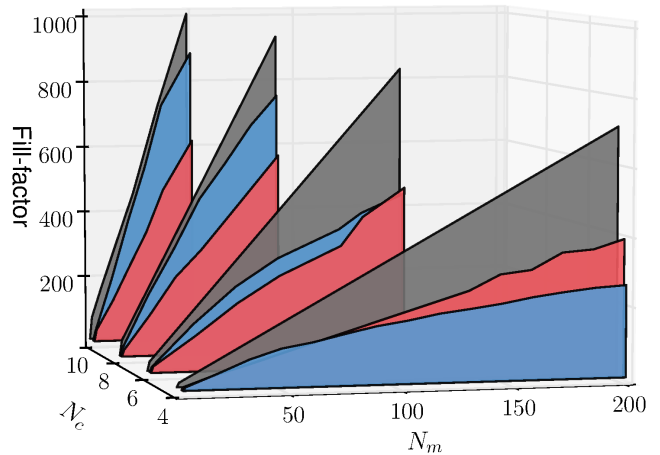}
\caption{(Color online) Fill-factors for the direct LU decomposition of the modified Liouvillian $\tilde{\bm{\mathcal{L}}}$ using natural (grey), RCM (red) and COLAMD (blue) matrix ordering as a function of Hilbert space dimensionality.  Here, the system parameters (in units of $\omega_{m}$) are: $\kappa=0.05$, $\Delta=-\kappa$, $g_{0}=3\kappa$,  $E=0.25$, $Q_{m}=10^{4}$, and $n_{\rm th}=31$.}
\label{fig:fig3}
\end{figure}

Prior to exploring the use of RCM and COLAMD reordering in iterative solutions, it is instructive to investigate the effect of these permutations, together with the natural matrix ordering, in the direct LU decomposition of $\tilde{\bm{\mathcal{L}}}$.  In Fig.~\ref{fig:fig3} we present the fill-factor $[\rm{NNZ}(L)+\rm{NNZ}(U)]/\rm{NNZ}(\tilde{\bm{\mathcal{L}}})$ for the direct LU decomposition of an optomechanical system with parameters similar to those for a superconducting circuit proposal that realizes the single-photon strong coupling regime, Ref.~\cite{rimberg:2014}.  
Simulations are performed in QuTiP \cite{qutip:2012, qutip:2013} using the SuperLU solver from SciPy \cite{scipy}.  Here, as in all other simulations presented in this work, we set the weighting factor $w$ in Eq.~(\ref{eq:linear}) to be the average of the diagonal elements in $\bm{\mathcal{L}}$.  This guarantees that the matrix elements corresponding to $\bm{\mathcal{T}}$ are not dropped during iLU factorization.  We see that in the naturally ordered modified Liouvillian $\tilde{\bm{\mathcal{L}}}$, the fill-in grows with a sharp linear dependence on the size of the underlying Hilbert space for the quadratically increasing bandwidth.  Applying either COLAMD or RCM reordering reduces the rate of growth, although the fill-in can still be three orders of magnitude or more greater than the original Liouvillian.  For a given $N_{c}$, RCM reordering outperforms COLAMD up until a given number of oscillator states, where the COLAMD fill-in rate becomes sub-linear.  As $N_{c}$ increases, this crossover value for $N_{m}$ rises, and RCM ordering produces a lower fill-in than COLAMD over most of the computationally tractable Hilbert space dimensions.  As the number of cavity states increases, the fill-in produced using COLAMD approaches that of the naturally ordered matrix, making the use of RCM essential in this regime. 
\begin{figure*}[t]
\includegraphics[width=17.2cm]{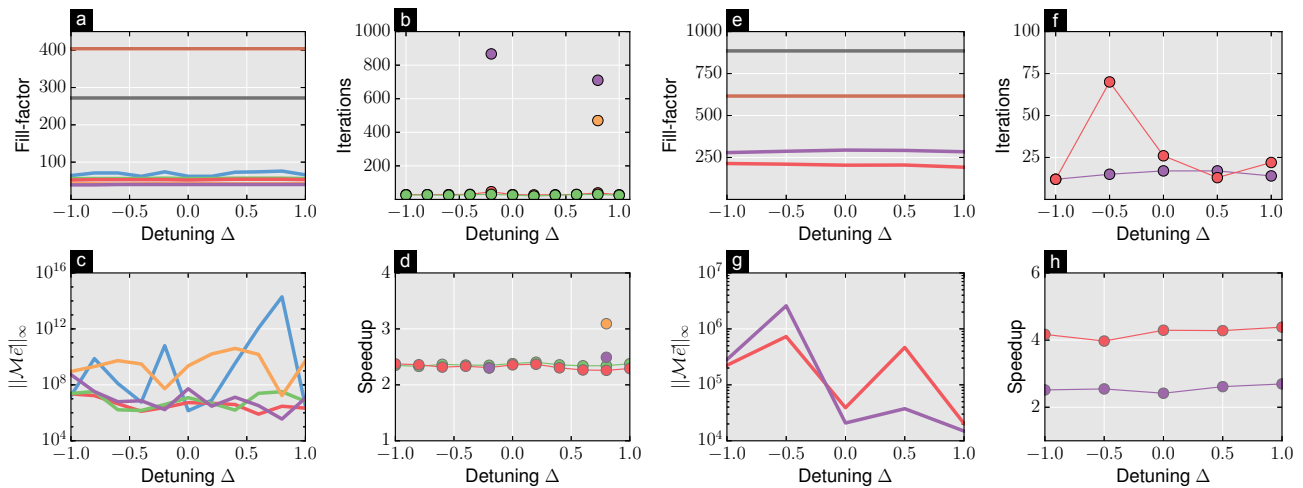}
\caption{(Color online) (a) Fill-factors as a function of detuning for direct LU factorization using RCM (brown) or COLAMD (grey) reordering, along with fill-factors for iLU preconditioners based on $\tilde{\bm{\mathcal{L}}}_{RCM}$ at $n_{\rm th}=0$ (blue), $n_{\rm th}=10^{-15}$ (green), and $n_{\rm th}=3$ (red).  Those based on the COLAMD ordering of $\tilde{\bm{\mathcal{L}}}$ at $n_{\rm th}=0$ (yellow) and $n_{\rm th}=3$ (purple) are also presented. (b) Number of iterations needed to reach machine precision for the iLU preconditioners constructed in (a).  Only those points where the solution converged within $1000$ iterations are displayed. (c) Log-plot of the estimated condition number $||\bm{\mathcal{M}}\vec{e}||_{\infty}$ for the iLU preconditioners given in (a). (d) Computational speedup of iterative solvers as measured against the runtime of direct LU factorization using COLAMD reordering.  Only those points where convergence was reached are displayed.  For (a-d), the number of cavity and mechanical oscillator states is $N_{c}=4$ and $N_{m}=200$, while the system parameters (in units of $\omega_{m}$) are:, $\kappa=0.2$,  $g_{0}=2.5\kappa$, $Q_{m}=10^{4}$, $E=0.1$. The iLU drop tolerance and allowed fill-in are $d=10^{-4}$ and $p=100$, respectively.  (e-h) Results of iterative simulations for the system parameters from Ref.~\cite{rimberg:2014}, and given in Fig.~\ref{fig:fig3}, for $N_{c}=10$ and $N_{m}=50$.  Here, the iLUTP parameters are set to $d=5\cdot 10^{-5}$ and $p=300$.  Only the direct LU (brown) and iterative (red) RCM results for $n_{\rm th}=31$, along with the corresponding COLAMD direct (grey) and iterative (purple) simulations are presented .}
\label{fig:fig4}
\end{figure*}

Turning to iterative simulations, we are interested in not only how matrix reordering affects these calculations, but also in how robust these methods are with respect to variations in system parameters that affect the underlying structure of $\tilde{\bm{\mathcal{L}}}$.  In particular, we focus on the detuning $\Delta$ and mechanical thermal occupation number, characterized by $n_{\rm th}$, both which are commonly set to zero, and thus eliminating some elements in the Liouvillian.  While it is customary to allow some error in the solution of iterative methods to aid in convergence, in order to rigorously test the performance gains of the iterative techniques presented here, we will investigate the memory and runtime requirements when converging to machine precession; the iterative answer is numerically identical to the solution vector found via direct LU decomposition.  This can be viewed as a worst-case scenario for iterative methods as it requires a high-precision, well conditioned $\bm{\mathcal{M}}$, and possibly a large number of iterations to reach this strict tolerance.  A convenient estimate for the condition number of the approximate inverse is given by  $||\bm{\mathcal{M}}\vec{e}||_{\infty}$, where $\vec{e}=(1,1,\dots)^{T}$ \cite{chow:1997} and $||\cdot||_{\infty}$ is the infinity-norm.  Although this measure gives only a lower bound on $||\bm{\mathcal{M}}||_{\infty}$, as we will show, it is an useful benchmark for the stability of $\bm{\mathcal{M}}$.  Having constructed an iLU preconditioner for a given drop tolerance $d$ and allowed fill-in $p$, we perform preconditioned iterations using the Restarted Generalized Minimum Residual (GMRES) solver \cite{saad:1986}.  The cost of each iteration grows as $\mathcal{O}(n^{2})$, where $n$ is the number of iterations, and therefore we restart the algorithm after 10 steps.  Increasing the number of iterations can improve convergence at the expense of memory usage.  Other methods suitable for non-symmetric matrices, such as the stabilized bi-conjugate gradient (Bi-CGSTAB) \cite{vorst:1992} and Loose GMRES (LGMRES) \cite{baker:2005} methods, were also investigated, but were found to have inferior convergence properties.  Finally, as sparse matrix-vector multiplication is limited by memory bandwidth \cite{im:2000}, we run all simulations in a serial, rather than parallel, fashion with the total computation time calculated from the average of three simulations.  This guarantees that memory bottlenecks do not effect the performance calculations, but is otherwise not necessary when using these techniques.  Indeed, the savings in memory consumption that iterative solvers offer, opens up the possibility for running multiple simulations in parallel.

In Fig.~\ref{fig:fig4}(a-d), we plot the results of iterative simulations of the optomechanical system from Ref.~\cite{nation:2013} with $N_{c}=4$ and $N_{m}=200$ using both RCM and COLAMD reordering of $\tilde{\bm{\mathcal{L}}}$ as a function of detuning.  In addition, we consider oscillator thermal environments at both zero and nonzero ($n_{\rm th}=3$) temperatures, the former being of theoretical interest.  Here, the iLUTP drop tolerance and allowed fill-in are set to $d=10^{-4}$ and $p=100$, respectively.  At this drop tolerance, in Fig.~\ref{fig:fig4}(a) we see that the fill-in for the iLU factorization is approximately a factor of five lower than the direct factorization using COLAMD ordering for preconditioners based on either RCM or COLAMD ordering, averaging $53$ and $39$, respectively for the nonzero thermal environment.  Fill-in values at zero temperature are slightly higher.  Although COLAMD gives lower fill-in, as seen in Fig.~\ref{fig:fig4}(b), only RCM reordering at nonzero temperatures converges to machine precision within 1000 iterations at all detunings.  RCM ordering at zero temperature completely fails, while COLAMD works only at a select few values for either zero or non-zero environments.  For those points that do not converge, the norm of the residual calculated from the approximate solution averaged $\sim10^{-2}$ instead of the requested $\le10^{-15}$.  This behavior can be understood by looking at the estimated condition number of $\bm{\mathcal{M}}$, Fig.~\ref{fig:fig4}(c), where it is seen that RCM reordering at nonzero temperature gives the lowest values, while this same reordering at zero temperature gives condition number estimates up to six orders of magnitude larger; the RCM iLU factors at zero temperature are too ill-conditioned to reach convergence.  For COLAMD reordering, the successful iterations occur at detunings where there is a reduction in the condition estimate, and therefore an increase in the stability of the preconditioner.  However, even for COLAMD, an oscillator environment at $n_{\rm th}=0$ gives rise to iLU factors that are very poorly conditioned.

The change in matrix structure resulting from assuming $n_{\rm th}=0$ makes the resulting Liouvillian operator less symmetric than it otherwise would be if a nonzero oscillator temperature were used; the structure of  $\tilde{\bm{\mathcal{L}}}+\tilde{\bm{\mathcal{L}}}^{T}$ is a poor representation of $\tilde{\bm{\mathcal{L}}}$ itself. This makes it difficult for permutations, such as RCM, that assume a symmetric structure to effectively reorder the matrix.  While it is possible to overcome this limitation using a small nonzero value for $n_{\rm th}$, to verify that it is the structure of $\tilde{\bm{\mathcal{L}}}$ that is responsible for this poor stability, we repeat our simulations using $n_{\rm th}=10^{-15}$.  This value is specifically chosen such that the structure of $\tilde{\bm{\mathcal{L}}}$ includes these thermal matrix elements, yet these terms are below the requested iLU drop tolerance ($d=10^{-4}$) and are not present in the preconditioner.  Moreover, this occupation number is so low as to not affect the final numerical answer;  the calculated density matrix is numerically identical to the $n_{\rm th}=0$ solution.  As seen in Fig.~\ref{fig:fig4}(c), this modification greatly increases the stability of preconditioning and convergence is once again achieved at all detunings.  As COLAMD is not sensitive to the same structural properties of $\tilde{\bm{\mathcal{L}}}$ as is RCM, this adjustment has no effect on COLAMD reordering.  The superior conditioning of iLU factors generated by RCM presented here, versus those using COLAMD, presented here is in line with previous studies on symmetric \cite{bridson:2000} and non-symmetric sparse matrices \cite{benzi:1999} where it was found that RCM reordering improves iLU conditioning at the expense of a marginal increase in fill-in.  Lastly, Fig.~\ref{fig:fig4}(d) shows the computational speed-up of the iterative solutions as measured against the runtime of direct factorization using COLAMD.  This measure is affected by both the iLU fill-in for creating the preconditioner, as well as the condition number of $\bm{\mathcal{M}}$ that determines the number of required iterations.  It is seen that, in addition to the factor of five reduction in fill-in, the RCM iterative solutions are over twice as fast when compared to direct LU methods.

The number of cavity states ($N_{c}=4$) considered in the previous simulations is valid only in the limit of a low-quality factor cavity or weakly driven system.  To examine the use of iterative methods in the high-quality cavity scenario, we repeat our simulations with the parameters given in Fig.~\ref{fig:fig3} for $N_{c}=10$.  As the direct LU fill-in is larger than in our previous simulations (see Fig.~\ref{fig:fig3}) we are limited to $N_{m}=50$ states for the mechanical oscillator.  The results when using an ILU drop tolerance $d=5\cdot 10^{-5}$ and fill-in limit of $p=300$ are presented in Fig.~\ref{fig:fig4}(e-h).  With a Hilbert space size less than half that of our previous example, the improvement in the conditioning of  $\tilde{\bm{\mathcal{L}}}$ allows for both RCM and COLAMD ordering to converge over the entire range of detunings, with average fill-in levels $\sim4.3$ and $\sim3$ times less than LU factorization using COLAMD, respectively.   Likewise, the condition estimates for $\bm{\mathcal{M}}$ in Fig.~\ref{fig:fig4}(g) are three-orders of magnitude lower than the lowest values presented in Fig.~\ref{fig:fig4}(c).  However, the poor performance of COLAMD in reducing fill-in at large $N_{c}$ allows iterative solutions using RCM to be twice as fast as those using COLAMD, and over four times faster than direct LU solutions.  By increasing the number of states for the mechanical oscillator, we have verified that preconditioners based on COLAMD reordering become unstable, while those using RCM still lead to convergence.

Our motivation in developing these iterative numerical techniques is to investigate the existence of nonclassical steady states  of mechanical motion in the single-photon strong-coupling regime where the presence of multiple limit-cycles prohibits time-dependent methods, and the dimensionality of the system Hilbert space is larger than what can be directly factored using typical computational resources.  In particular, we wish to verify the existence of such states using experimentally feasible parameters for the superconducting cavity-Cooper pair transistor (c-CPT) scheme from Ref.~\cite{rimberg:2014}, which is predicted to be well within the single-photon strong-coupling regime, $g^{2}_{0}/\kappa\omega_{m}\sim 3$.  However, for a realizable oscillator frequency of $\omega_{m}=2\pi\times 10\rm{MHz}$, the average resonator bath occupation number is $n_{\rm{th}}\sim 20$ at $10\rm{mK}$, and we must find parameters for which nonclassical signatures such as negativity in the oscillator Wigner function and/or sub-Poissonian statistics of the individual oscillator limit-cycles survive at these experimentally accessible temperatures.  

\begin{figure}[t]
\includegraphics[width=7.0cm]{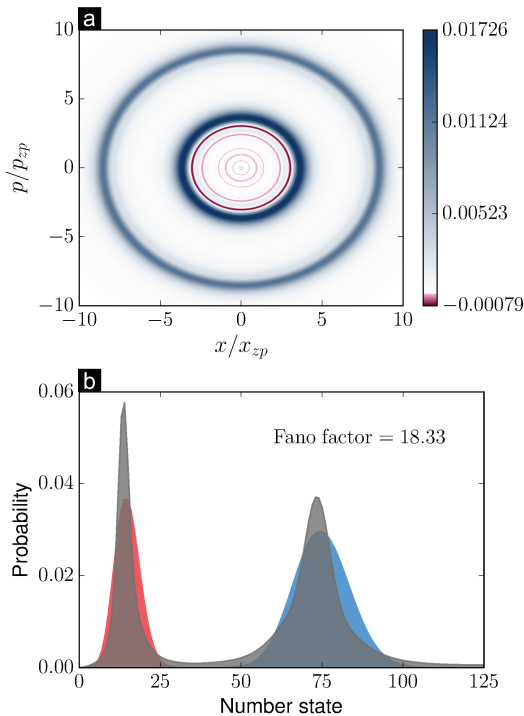}
\caption{(Color online) (a) Wigner distribution for the mechanical oscillator density matrix at (in units of $\omega_{m}$): $E=0.45$,  $\kappa=0.05$, $\Delta=-2\kappa$, $g_{0}=0.32$, and $n_{\rm th}=20$.  Here, $x_{\rm zp}$ and $p_{\rm zp}$ are the zero-point amplitudes of the oscillator's position and momentum, respectively. (b) Phonon number distribution (grey) for the oscillator given in (a) along with fitted coherent states for the lower (red) and upper (blue) limit-cycles.  The total Fano factor for the distribution is also given. Simulation parameters are: $N_{c}=5$, $N_{m}=160$, $d=10^{-4}$, and $p=200$.  Only the first $125$ oscillator Fock states are displayed for better limit-cycle visibility.}
\label{fig:fig5}
\end{figure}

From an earlier investigation \cite{nation:2013} it is known that the strongest nonclassical Wigner functions, as measured by the proportion of negative Wigner area, occur at, or just above, the renormalized cavity resonance frequency $\omega_{c}'=\omega_{c}-g_{0}^{2}/\omega_{m}$, and that the survival of these features at nonzero temperatures requires being well within the single-photon strong-coupling regime.  Given this information, as an illustrative example, we solve for the steady-state density matrix at $\Delta/\omega_{m}=-2\kappa/\omega_{m}=-0.1, g_{0}/\omega_{m}=0.32$ ($g^{2}_{0}/\kappa\omega_{m}\sim2$) using a pump power $E/\omega_{m}=0.45$ and present the results in Fig.~\ref{fig:fig5}.

The negativity in the Wigner function for the mechanical resonator density matrix seen in Fig.~\ref{fig:fig5}a is a signature that the mechanical oscillator is in a nonclassical state.  This state consists of two partially overlapping limit-cycles, Fig.~\ref{fig:fig5}b, with an overall phonon probability distribution that gives a Fano factor, $F=\langle(\Delta \hat{N}_{b})^{2}\rangle/\langle \hat{N}_{b}\rangle$, much larger than one, indicating super-Poissonian ($F>1$) statistics.  However, it is well known that negativity in the oscillator Wigner function arrises from the sub-Poissonian ($F<1$) statistics of individual limit-cycles, when the narrowed phonon number distribution is sufficiently close to that of a mechanical Fock state \cite{rodrigues:2010,qian:2012,nation:2013,lorch:2014}.  To qualitatively show that each limit-cycle still possesses sub-Poissonian features, in Fig.~\ref{fig:fig5}b we fit each effective limit-cycle with a coherent state, $F=1$, of the same mean phonon number, and normalized to the probability of each limit-cycle.  It is seen that the phonon number distribution for each limit-cycle is narrower than that of the corresponding fitted coherent state, in marked contrast to the tails of the distributions that are broadened due to the partial merging of the two limit-cycles and effects from the oscillator thermal bath.  As a steady state of the system dynamics, this persistent quantum state can be repeatedly measured using quantum state tomography \cite{lvovsky:2009} without loss of fidelity.  With the parameters used here, an accurate simulation of the system requires $N_{c}=5$ and $N_{m}=160$, which is  outside the direct LU factorization capabilities of our computational hardware, and must be solved using iterative methods.  While Fig.~\ref{fig:fig5} shows the promise of the c-CPT device with respect to the generation of nonclassical mechanical states of motion, the results presented here can likely be improved upon with a more detailed search of parameter space.  Such a search is aided by the reduced memory footprint of iterative methods, allowing for several simulations to be run in parallel, thus greatly reducing the computation time.

All of the examples presented here highlight the performance of iterative methods under the strict condition of convergence to machine precision.  Under more typical usage scenarios, where a given amount of error is allowed in the computed density matrix, one can selectively tune the level of fill-in in the iLU factors by varying the iLUTP drop tolerance.  Specifying an acceptable solution error, as measured by the norm of the residual vector, it is possible to lower the iLU fill-in at the expense of an increase in the number of iterations needed for convergence.  For an optimal drop tolerance $d$, it is possible to reduce the memory requirements of factorization while maintaining enough accuracy in the iLU factors so as to converge to the relaxed solution tolerance in a reduced amount of time.  Therefore, it is possible to improve upon the performance of the methods presented here with respect to both memory and runtime measures.  Note however that the stability of $\bm{\mathcal{M}}$ also depends on the drop tolerance in a non-trivial manner, and therefore only select values for the drop tolerance will lead to convergence. 
 
\section{Conclusion}\label{sec:conclusion}
We have shown that RCM reordering of the Liouvillian super operator for quantum optomechanical systems gives stable iLU preconditioners necessary for iterative solutions to the steady state density matrix.  The combination of low fill-in and condition number, allows these iterative techniques to outperform direct solution methods in terms of both memory consumption and computation runtime.  The results presented here are limited only by the memory constraints of direct LU factorization, and there is nothing that prohibits these relative performance benchmarks from being an order of magnitude, or higher, as the dimensionality of the Hilbert space is increased.  Unlike other matrix permutation methods, the iLU factors generated by RCM reordering are stable over the entire parameter space, provided that one correctly adjusts for the special case of a zero temperature oscillator environment.  Although we have only examined the standard optomechanical Hamiltonian, the benefits of RCM reordering apply equally well to variants of  Eq.~(\ref{eq:hamiltonian}) provided that the Hamiltonian remains time-independent.  In particular, RCM reordering is well suited for quadratic ``membrane in the middle" optomechanical systems where the coupling term takes the form $(\hat{b}+\hat{b}^{\dag})^{2}\hat{a}^{\dag}\hat{a}$ \cite{thompson:2008,sankey:2010}, as well as additional anharmonic oscillator terms \cite{rips:2012}.   In addition, since any symmetric permutation is a similarity transformation, RCM reordering can also be applied in iterative eigenvalue solvers.  

The extension of this reordering scheme to other classes of Hamiltonians is less straightforward.  While RCM reordering of a symmetric matrix is guaranteed to reduce the bandwidth and profile, or at least do no worse \cite{george:1981}, the need to operate on $\tilde{\bm{\mathcal{L}}}+\tilde{\bm{\mathcal{L}}}^{T}$ rather than $\tilde{\bm{\mathcal{L}}}$ itself invalidates this assurance.  For example, permuting a Jaynes-Cummings system \cite{jaynes:1963} using RCM reordering leads to an increase in both bandwidth and profile, and a subsequent growth in memory and runtime requirements.  However, this is not to say that matrix permutation methods can not be employed for these systems.  Indeed, for system Liouvillians, like the Jaynes-Cumming model, where zeros can occur along the main diagonal, non-symmetric reorderings such as maximum cardinality bipartite matching methods \cite{duff:2011}, or weighted variants thereof, that permute the diagonal to be zero-free must be used before even standard reorderings such as COLAMD can be employed.   An, as yet unknown, symmetric permutation that reduces the bandwidth and profile of the resulting matrix can then be applied \footnote{Symmetric permutations do not change the values along the main diagonal, although they can be reordered.}.  

Finding a successful permutation strategy for a given system Hamiltonian can be a daunting task given the plethora of reorderings available, the non-symmetric structure of the Liouvillian, and the need to empirically test each one.  However, as we have shown, the stability of the approximate inverse is the key requirement when using iterative solvers.  As such, theoretical \cite{bridson:2000}, and numerical \cite{benzi:1999} evidence points to permutations based on reversed breadth-first (such as RCM) or depth-first search methods as possible candidates.  Once found, a successful permutation scheme unlocks the possibility of using iterative solution methods, whihc appear to be the only methods available when solving very large-scale linear systems.  Although RCM reordering itself is a serial algorithm, the methods presented here are scalable to parallel and distributed computing architectures \cite{saad:2003,li:2003, amestoy:2000} and, with a quantum computer out of reach for the foreseeable future, represent the best available solution method for the steady state density matrix of optomechanical and other time-independent open quantum systems.

All of the tools presented in this work can be found in the open-source QuTiP framework \cite{qutipweb}.  In addition, the Python scripts used for generating the results presented here are available as ancillary files with the arXiv version of the manuscript.

\section*{Acknowledgements}
PDN was supported by startup funding from Korea University.  MPB was supported by the NSF under Grant No. DMR-1104790.  AJR was supported the NSF under Grant No. DMR-1104821, by AFOSR/DARPA agreement FA8750-12-2-0339, and by the ARO under Contract No. W911NF-13-1-0377.

\bibliography{refs}

\end{document}